\documentclass{elsart}

\usepackage{amssymb,amsfonts,amsmath}
\usepackage[dvips]{epsfig}

\newcommand{\bs}{\boldsymbol}

\begin{document}

\begin{frontmatter}
  
  \title{Low-temperature antihydrogen-atom scattering }

  \author{S Jonsell} \ead{jonsell@tp.umu.se} \address{ Department of
    Physics, Ume{\aa} University, SE-90187 Ume{\aa}, Sweden }

\begin{abstract}
  A simple method to include the strong force in atom--antiatom
  scattering is presented. It is based on the strong-force scattering
  length between the nucleon and antinucleon. Using this method
  elastic and annihilation cross sections are calculated for
  hydrogen--antihydrogen and helium--antihydrogen scattering. The
  results are compared to first-order perturbation theory using a
  pseudo potential. The pseudo-potential approach works fairly well
  for hydrogen--antihydrogen scattering, but fails for
  helium-antihydrogen scattering where strong-force effects are more
  prominent.
\end{abstract}


\end{frontmatter}

Antimatter in contact with matter causes annihilation -- a violent
process where mass is converted to energy. This is well known, even
among people who never have studied physics antimatter annihilation
seems to stimulate the imagination. In popular films and literature
antimatter is used as e.g.\ an energy source, fuel for space ships, or
a weapon.  However, since the energy required to create antimatter
outweighs the energy created by annihilation by many orders of
magnitude, neither of these applications are possible in reality.
More realistically, antimatter may have applications in medicine, such
as antiproton treatment of cancer.

The primary goal for antimatter research is not futuristic
applications, but to uncover the symmetries (or possible asymmetries)
of the fundamental laws of physics. One could expect that the matter
and antimatter contents of the universe would be symmetric, but so far
observations indicate that at least in the neighborhood of our galaxy
there is only matter.  Today the coexistence of matter and antimatter
is ruled out on scales $\lesssim 20$~Mpc \cite{coh97}. The reason for
this matter--antimatter asymmetry could be a violation of the CPT
theorem. This theorem, and other matter--antimatter symmetries, could
be tested through high-precision measurements on antihydrogen.  Two
experiments at CERN, ATHENA and ATRAP, have succeeded in creating
antihydrogen atoms, but much work remains before CPT tests can be
carried out \cite{antih}.

Despite the interests in antimatter of both scientists and the general
public, relatively little is known about the interaction between atoms
made of antimatter and ordinary atoms. So far no experimental data
exist, and theoretical work has only just started. Atom-antiatom
scattering has practical implications for antihydrogen experiments,
e.g.\ unwanted scattering of antiatoms on impurities or on the walls
of the experiment, or scattering on ultracold atoms introduced on
purpose to achieve sympathetic cooling of antimatter. Atom--antiatom
scattering is also interesting in its own right, being a new
fundamental process involving both atomic and low-energy nuclear
physics.

A number of inelastic processes are possible in atom-antiatom
scattering: nucleus-antinucleus annihilation, electron-positron
annihilation, rearrangement processes (such as formation of
positronium), and even formation of metastable ``molecules''.  The
most important of these are nucleus-antinucleus annihilation and
rearrangement.  Rearrangement processes will also eventually lead to
annihilation, but on a time scale much longer than the time required
for the final-state fragments to separate. Hence, it can be regarded
as a process distinct from the ``direct'' annihilation during the
collision. Alternatively, rearrangement can be regarded as a resonance
in the annihilation channel.  In this paper I will focus on the direct
annihilation process, while rearrangement is discussed extensively
elsewhere in these proceedings \cite{arm06}.

At large internuclear separations $R$ the wave function of the
atom-antiatom system with relative angular momentum $l$ is given by
\begin{equation}
  \label{eq:Rinf}
  \Phi({\bs R},{\bs r}_i)\rightarrow N \frac{1}{R}\sin(k_i
  R-l\pi/2+\delta_l)Y_{lm}(\Omega_R)\,\psi_{\bar{\rm H}}({\bs r}_1)
  \phi_{\text{atom}}({\bs r}_{i>1}),
\end{equation}
where $\psi_{\bar{\rm H}}$ and $\phi_{\text{atom}}$ are the atomic
wave functions of antihydrogen and the atom, ${\bs r}_i$ denotes the
coordinates of all leptons, and $\hbar k_i$ is the relative momentum
in the initial state. Both elastic and annihilation cross sections are
given by the complex phase shift $\delta_l$ \cite{lan77},
\begin{align} \label{eq:sigmael}
  \sigma^{\rm el}& =\frac{\pi}{k_i^2}(2l+1)\left|1-e^{2i\delta_l}
  \right|^2,\\ \label{eq:sigmainel} \sigma^a&
  =\frac{\pi}{k_i^2}(2l+1)\left(1-e^{-4\text{Im}\delta_l}\right).
\end{align}
At low temperatures ($\lesssim 1$~K) only the $l=0$ partial wave
contributes, and the phase shift has the low-energy form
\begin{equation}
\label{eq:a}
  \lim_{k_i\rightarrow0}\tan\delta_0=-k_i a,
\end{equation}
where $a=\alpha-i\beta$ is the complex scattering length. Inserting
(\ref{eq:a}) into (\ref{eq:sigmael}) and (\ref{eq:sigmainel}) gives
the low-energy form of the cross sections
\begin{align}
  \sigma^{\rm el}& =4\pi\left(\alpha^2+\beta^2\right),\\
  \sigma^a& =\frac{4\pi}{k_i}\beta.
\end{align}
We first note that the elastic cross section is constant at low
energies, while the annihilation cross section diverges as $k_i^{-1}$.
Thus, below a certain energy annihilation dominates. Second, we note
that $\sigma^{\rm el}$ depends on both $\alpha$ and $\beta$. This
means that a non-zero imaginary part $\beta$ of the scattering length,
arising from the inclusion of the strong nuclear force, will not only
give rise to annihilation, but will also significantly modify the
elastic cross section. Only if the annihilation cross section is
small, $\beta\ll |\alpha|$, this effect may be ignored.

The simplest way to calculate the annihilation cross section is to use
a pseudo potential
\begin{equation}
  \label{eq:pseudo}
  V_a({\bs R}) =A \delta ({\bs R}).
\end{equation}
This is a contact interaction with strength given by the constant $A$.
This constant can be determined from experimental data on the
nucleus-antinucleus system, or it can be calculated from model
potentials. Using the pseudo potential in first-order perturbation
theory the annihilation cross section becomes
\begin{equation}
  \label{eq:sigmaa}
  \sigma^a=\frac{(2\pi)^3}{k_i^2}A \lim_{R\rightarrow0}\int\left| \Phi({\bs R},{\bs
      r}_i)/R\right|^2 d{\bs r}_i.
\end{equation}
For the proton-antiproton system $A^{p\bar{p}}=1.7\times
10^{-7}$~a.u.=$6.8\times 10^{-37}$~eVm$^3$, which was determined from
the width 1130~eV of the $1s$ state of protonium \cite{bat96}. For the
alpha particle-antiproton system no data on the $1s$ state are
available, instead the value $A^{\alpha\bar{p}}=3.4\times
10^{-7}$~a.u.=$1.4\times 10^{-36}$~eVm$^3$ was determined from a
combination of low-energy annihilation data, the energy shift of the
$2p$ state, and the width of the $2p$ and $3d$ state \cite{bat01}.
Both values represent an average over different spin states.

The pseudo-potential approach, being based on first-order perturbation
theory, does not take into account the modification of the initial
channel due to the strong force.  Even if annihilation cannot be
treated perturbatively one may still use the zero-range approximation
of the strong force, since the atomic nucleus is very small on an
atomic scale. The detailed shape of the strong-force potential will
therefore not matter, only its strength as parameterized by a
short-range strong-force phase shift $\delta_{sf}$. Furthermore,
atomic energies (eV) are very small compared to typical nuclear
energies (MeV). The strong-force phase shift can therefore be
parameterized by a strong-force scattering length $a_{sf}$.  This
scattering length is complex, with the imaginary part representing
annihilation.

The role of the strong-force scattering length can be understood by
looking at the boundary condition of the wave function at short
internuclear separations, $R\rightarrow0$. In the absence of the
strong force, the wave function has a short-distance form proportional
to the regular Coulomb function for zero angular momentum $F_0(kR)$.
Even when the strong force is added the wave function is Coulombic for
$R_0\leq R \ll 1a_0$, where $R_0$ is just outside the range of the
strong force. But now, since the Coulombic solution is not extended
all the way to $R=0$, also the irregular Coulomb function $G_0(kR)$ is
allowed. A general eigensolution to the zero-angular momentum Coulomb
problem is an arbitrary linear combination of $F_0(kR)$ and $G_0(kR)$.
For atom-antiatom scattering in the presence of the strong nuclear
force the appropriate short-range form of the scattering function is
given by a particular solution determined by the strong-force phase
shift $\delta_{sf}$
\begin{equation}
  \label{eq:Phishort}
  \Phi({\bs R},{\bs r}_i)=\frac{
    F_0(kR)+\tan\delta_{sf}G_0(kR)}{F_0(kR_0)+\tan\delta_{sf}G_0(kR_0)} 
\varphi({\bs r}_i). 
\end{equation}
Here $\varphi({\bs r}_i)$ is the wave function of the leptons in the
combined electric field of the overlapping nucleus and antinucleus.
The wave vector $k$ is related to $k_i$ as
$\hbar^2k^2/(2\mu)=\hbar^2k_i^2/(2\mu)-E_f$, where $E_f$ is the
leptonic energy corresponding to $\varphi({\bs r}_i)$.  Since $R_0$ is
very small on an atomic scale, the boundary condition of $\Phi$ at
$R_0$ may be expressed in terms of the short-range expansions of the
Coulomb functions \cite{abr},
\begin{align}
  F_0(kR) & \rightarrow C_0(\eta) k R ,\\
  G_0(kR) & \rightarrow C_0^{-1}(\eta)\left\{1-2\eta k
    R\left[\ln(2\eta kR)+h(\eta)+2\gamma-1\right]\right\}.
\end{align}
Here the strength of the Coulomb interaction is given by
$\eta=1/(b_{\mu}k)$, where $b_{\mu}=a_0m_e/\mu$ is the mass-scaled
Bohr radius. The normalization constant is
$C_0(\eta)=\sqrt{2\pi\eta/[1-\exp(-2 \pi \eta)]}$, $\gamma\simeq
0.5772\dots$ is the Euler constant, and $h$ can be expressed in terms
of the digamma function $\Psi$ as
\begin{equation}
  h(\eta)=\left[\Psi(-i\eta)+\Psi(i\eta)\right]/2-\ln\eta .
\end{equation}
One then finds that \cite{tru61}
\begin{equation}
  \label{eq:cotdsi}
  \frac{C_0^2(\eta)}{\eta}\cot\delta_{sf}-2h(\eta)=-\frac{ b_{\mu}}{a_{sf}},
\end{equation}
where the strong-force scattering length is
\begin{equation}
  \label{eq:asi}
  \frac{1}{a_{sf}}=-\frac{2\pi}{b_{\mu}}\lim_{k\rightarrow0}\cot\delta_{sf}.
\end{equation}

The wave function $\Phi$ may be expressed in terms of a regular and an
irregular solution as
\begin{align}
  \label{eq:Phi2}
  \Phi({\bs R},{\bs r}_i)=&-\frac{b_\mu^2/(a_{sf}R_0)
    F_0(kR_0)}{\left[2h(\eta)-b_\mu/a_{sf}\right]
    F_0(kR_0)+C_0(\eta)^2/\eta G_0(kR_0)}\Phi^{\rm reg}({\bs R},{\bs
    r}_i) \nonumber\\ & + \frac{2h(\eta)F_0(kR_0)+C_0^2(\eta)/\eta
    G_0(kR_0)}{\left[2h(\eta)-b_\mu/a_{sf}\right]
    F_0(kR_0)+C_0(\eta)^2/\eta G_0(kR_0)} \Phi^{\rm irreg}({\bs
    R},{\bs r}_i).
\end{align}
The regular and irregular solutions used are defined by their
short-range forms
\begin{align}
  \label{eq:Psireg}
  \lim_{R\rightarrow0}\Phi^{\text{reg}}({\bs R},{\bs r}_i)&=
  \frac{R_0}{b_\mu F_0(kR_0)} F_0(kR)\varphi({\bs r}_i)\simeq
  \frac{R}{b_\mu}\varphi({\bs r}_i), \\
  \label{eq:Psiirreg}
  \lim_{R\rightarrow0} \Phi^{\text{irreg}}({\bs R},{\bs r}_i)&=
  \frac{1}{\frac{2\eta h(\eta)}{C_0(\eta)^2} F_0(kR_0)+G_0(kR_0)}
  \left[\frac{2\eta h(\eta)}{C_0(\eta)^2} F_0(kR)+ G_0(kR)
  \right]\varphi({\bs r}_i) \nonumber \\ & \simeq
  \left\{1-\frac{2R}{b_\mu}\left[\ln\left(\frac{2R}{b_\mu}\right)+2\gamma-1\right]
  \right\}\varphi({\bs r}_i).
\end{align}
These forms have the advantage that they are independent of $R_0$,
provided that $R_0$ is small enough.  The normalization factors ensure
that $\Phi^{\rm reg}$ and $\Phi^{\rm irreg}$ are real functions, even
if $k$ is not. For large internuclear separations
\begin{align}
  \label{eq:Psireg2}
  \lim_{R\rightarrow\infty}\Phi^{\text{reg}}({\bs R},{\bs r}_i)&=
  N_{\rm reg}\frac{1}{R}\sin(k_iR+\delta_{\rm reg})\psi_{\bar{\rm
      H}}({\bs r}_1)
  \phi_{\text{atom}}({\bs r}_{i>1}),\\
  \label{eq:Psiirreg2}
  \lim_{R\rightarrow\infty} \Phi^{\text{irreg}}({\bs R},{\bs r}_i)&=
  N_{\rm irreg}\frac{1}{R}\sin(k_iR+\delta_{\rm irreg})\psi_{\bar{\rm
      H}}({\bs r}_1) \phi_{\text{atom}}({\bs r}_{i>1}),
\end{align}
where both normalization factors and phase shifts are real. Inserting
(\ref{eq:Psireg2}) and (\ref{eq:Psiirreg2}) into (\ref{eq:Phi2}), and
using (\ref{eq:Rinf}), the complex phase shift $\delta_0$ for
atom-antiatom scattering including the strong nuclear force is found
to be
\begin{equation}
  \label{eq:delta}
  \cot\delta_0=\frac{ N_{\rm
      reg}\cos\delta_{\rm reg}- a_{sf}/b_\mu N_{\rm irreg}\cos\delta_{\rm
      irreg}}{N_{\rm reg}\sin\delta_{\rm reg}-a_{sf}/b_\mu N_{\rm irreg}\sin\delta_{\rm irreg}} .
\end{equation}
Here the approximation $kR_0\rightarrow0$ has been used.  With this
equation we arrive at a very simple way of calculating the
strong-force effects beyond the pseudo potential in first-order
perturbation theory. In addition to the usual regular solution
(\ref{eq:Psireg}), an additional irregular solution given by the
boundary condition (\ref{eq:Psiirreg}) has to be calculated. The
asymptotic forms of the two solutions is then weighted according to
(\ref{eq:delta}) using the strong-force scattering length.

The scattering length method has several advantages. It is very easy
to implement numerically. It makes no assumptions about the detailed
form of the strong-force potential, and is hence model independent.
Alternatively one may directly use Eq.\ (\ref{eq:Phishort}), in
combination with Eq.\ (\ref{eq:cotdsi}), as a short-range boundary
condition, and integrate towards $R=\infty$. This was the procedure in
Refs.\ \cite{jon05,jon04b,jon04}. The present formulation in terms of
Eq.\ (\ref{eq:delta}) has the advantage that it clearly brings out the
relation between the atom-antiatom phase shift $\delta_0$ and the
strong-force scattering length $a_{sf}$, in terms of a minimal number
of parameters. The parameters $N_{\rm reg}$, $N_{\rm irreg}$,
$\delta_{\rm reg}$, and $\delta_{\rm irreg}$ are independent of the
strong force. Once they have been determined from an atom-antiatom
calculation, it is a trivial task to calculate $\delta_0$ for a given
strong-force scattering length.
 
\begin{figure}[htbp]
  \includegraphics*[width=12cm]{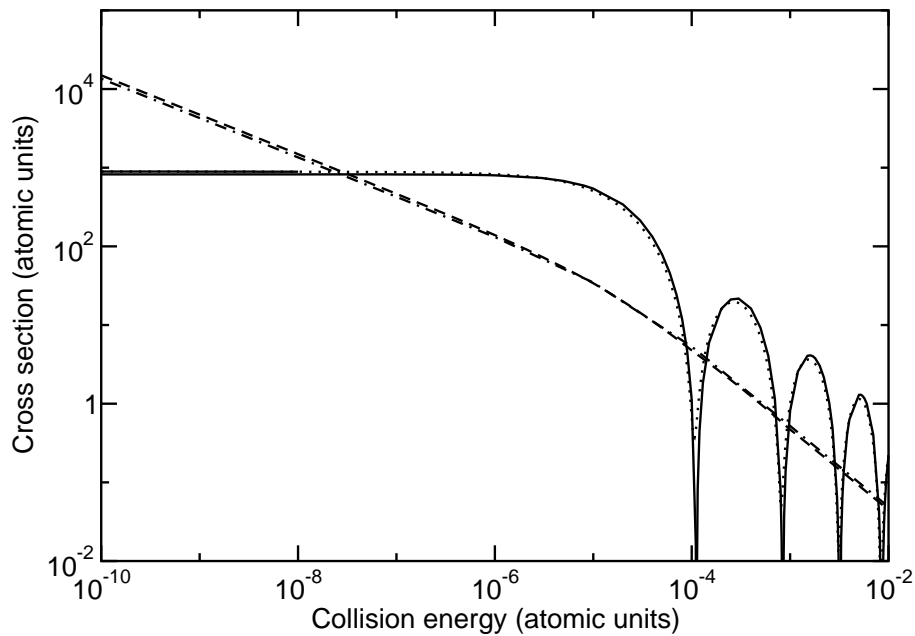}
  \caption{Cross sections for H--$\bar{\rm H}$ scattering; (solid)
    elastic cross section including the strong force, (dotted) elastic
    cross section without the strong force, (dashed) annihilation
    using the scattering length approach (dash-dotted) annihilation
    using pseudo potential. Only the $s$-wave part of the elastic
    cross section is shown.}
  \label{fig:HaH}
\end{figure}

\begin{figure}[htbp]
  \includegraphics*[width=12cm]{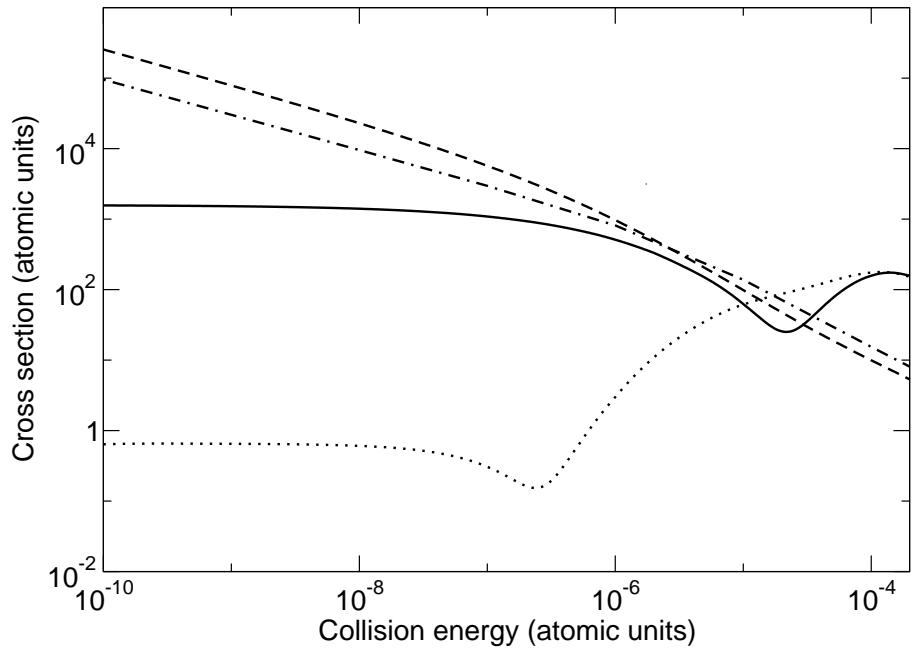}
  \caption{Cross sections for He--$\bar{\rm H}$ scattering; (solid)
    elastic cross section including the strong force, (dotted) elastic
    cross section without the strong force, (dashed) annihilation
    using the scattering length approach (dash-dotted) annihilation
    using pseudo potential.}
  \label{fig:HeaH}
\end{figure}

Using this method elastic and annihilation cross sections have been
calculated for H--$\bar{\rm H}$ and He--$\bar{\rm H}$. The
strong-force scattering lengths used were $a_{sf}^{p\bar
  p}=(0.844-0.698i)$~fm, and $a_{sf}^{\alpha\bar
  p}=(1.851-0.630i)$~fm, determined from the same data as the
annihilation constants $A^{p\bar p}$ and $A^{\alpha \bar p}$ above.
Once again, these are averages over different spin states.  The
H--$\bar{\rm H}$ potential was taken from Ref.\ \cite{str02b} and the
He--$\bar{\rm H}$ potential from Ref.\ \cite{str05}.  In figures
\ref{fig:HaH} and \ref{fig:HeaH} we compare the pseudo-potential and
the scattering-length approaches for H--$\bar{\rm H}$ and
He--$\bar{\rm H}$ respectively.

For H--$\bar{\rm H}$ the pseudo-potential approach works quite well.
As has been reported earlier, the low-energy annihilation cross
section using the pseudo-potential approach is
$0.14/\sqrt{\epsilon_i}\,a_0^2$ \cite{jon01}, while the scattering
length approach gives $0.15/\sqrt{\epsilon_i}\,a_0^2$ \cite{jon04b}
($\epsilon_i$ is the collision energy). This is also in good agreement
with a calculation by Armour et al. \cite{arm05}, using the optical
potential of Kohno and Weise \cite{koh86}, who obtained
$0.12/\sqrt{\epsilon_i}\,a_0^2$ for the triplet state and
$0.15/\sqrt{\epsilon_i}\,a_0^2$ for the singlet state. The elastic
cross section changes from $771\, a_0^2$ to $892\, a_0^2$ when the
strong-force effects are included. The change obtained by Armour et
al.\ \cite{arm05} was from $788\, a_0^2$ to  $920\, a_0^2$ for the
triplet and $872\,a_0^2$ for the singlet. The scattering length is
$a=(8.4-0.51i)\,a_0$, so indeed $\beta \ll |\alpha|$ as required for
the validity of the pseudo-potential approach.

For He--$\bar{\rm H}$ scattering the situation is quite different. The
strong force has a drastic influence on the elastic cross section.
This is in part because the real part of the scattering length happens
to have an untypically low value $\alpha=-0.23\,a_0$ (the typical
scale of the scattering length is $\sim 5\,a_0$). The elastic cross
section without the strong force is therefore only $0.65\,a_0$, but
changes drastically to $1700\,a_0^2$ when the strong force is
included.  This is reflected by the scattering length $a=(2.2-
11.4i)\,a_0$ , which has an imaginary part much larger than the real
part. The annihilation cross section changes from
$0.95/\sqrt{\epsilon_i}\,a_0^2$ in the pseudo-potential approach to
$2.62/\sqrt{\epsilon_i}\,a_0^2$ in the scattering-length approach.

These results replace those in Ref.\ \cite{jon04} which were based on
the potential in Ref.\ \cite{str02}. This potential had been
calculated at fewer $R$-values, giving interpolation problems in the
cross section calculations. Rather surprisingly, the effect of this
refinement of the potential on both elastic and annihilation cross
sections was quite large. The scattering length reported in Ref.\ 
\cite{jon04} was $a=(-7.69-3.80i)\,a_0$, giving an annihilation cross
section $0.88/\sqrt{\epsilon_i}\,a_0^2$.  The increase of the
calculated annihilation cross section with a factor three can almost
 entirely be
attributed to the range $R\lesssim1.5a_0$ of the potential. The
difference between the old and the new potential is very small, at most about
0.007~a.u. (at $R=0.8a_0$), showing the extreme sensitivity of the
scattering length to the details of the potential. Although this is a
small correction, it is still 20 times larger than the maximum of the
adiabatic correction \cite{str05}, which on the contrary only has a
small impact on the scattering length.

The validity of the scattering length approach has been tested for the
H--$\bar{\rm H}$ system using a model potential with the form
\cite{bat01}
\begin{equation}
  \label{eq:Vopt}
  V_{\rm
    opt}(R)=-\frac{2\pi\hbar^2}{\mu}\left(1+\frac{A-1}{A}\frac{\mu}{m}\right) b_0 \rho(R).
\end{equation}
Here $A$ is the nuclear mass number, $b_0$ a complex parameter, $\mu$
the antiproton-nucleon reduced mass, $m$ the mass of the nucleon, and
$\rho(R)$ the nuclear mass density, which was taken to have a Gaussian
form with width $r_G$. Three different widths were used, $r_G=1.0,
1.5, 2.0$~fm, and for each width the parameter $b_0$ was adjusted
until the potential reproduced the scattering length above. All three
potentials thus obtained gave results identical to the
scattering-length approach, to within three significant digits
\cite{jon05}. This clearly shows that the shape of the potential does
not matter, and all its relevant properties can be summarized by the
complex strong-force scattering length $a_{sf}$.

In conclusion I have developed a simple method to take the strong
force into account in situations when the pseudo-potential method
cannot be applied. This is the case in He--$\bar{\rm H}$ scattering,
where annihilation is fast, and the elastic cross section is
drastically modified by the strong force. In the future this method
may be used to investigate scaling properties of elastic and
annihilation cross sections for scattering of antihydrogen on heavier
atoms.

\end{document}